\documentstyle[12pt]{article}

\begin{document}
\topmargin 0pt
\oddsidemargin 7mm
\headheight 0pt
\topskip 0mm

\addtolength{\baselineskip}{0.40\baselineskip}

\hfill SOGANG-HEP 260/99

\hfill August 9 , 1999

\begin{center}

\vspace{36pt}
{\large \bf Consistent Superalgebraic Truncations from {\it D}=5,
{\it N}=5 Supergravity}

\end{center}

\vspace{36pt}

\begin{center}

Chang-Ho Kim$^a$

{\it Department of Physics and Basic Science Research Institute, Seonam University, \\
Namwon, Chonbuk 590-711, Korea}

\vspace{12pt}

Young-Jai Park$^b$

{\it Department of Physics and Basic Science Research Institute, Sogang University, \\
C.P.O. Box 1142, Seoul 100-611, Korea}

\end{center}

\vspace{1cm}
\begin{center}
{\bf ABSTRACT}
\end{center}

We study a novel five-dimensional, {\it N}=5 supergravity in the context of Lie superalgebra SU(5/2). The possible successive superalgebraic truncations from {\it N}=5 theory to the lower supersymmetric {\it N}=4,3,2, and 1 supergravity theories are systematically analyzed as a sub-superalgebraic chain of SU(5/2)$\supset$ SU(4/2) $\supset$ SU(3/2) $\supset$ SU(2/2) $\supset$ SU(1/2) by using the Kac-Dynkin weight techniques.   

\vspace{2cm}


\vspace{12pt}

\noindent

\vspace{12pt}

\vfill
%\vspace{2cm}
\hrule
\vspace{0.5cm}
\hspace{-0.6cm}$^a$ E-mail address : chkim@tiger.seonam.ac.kr \\
$^b$ E-mail address : yjpark@ccs.sogang.ac.kr

\newpage
\begin{center}
{\large \bf I. Introduction}
\end{center}

There have been considerable interests in superalgebras which are relevant to many supersymmetric theories.$^{1,2}$
Supersymmetric extensions of Poincar\'{e} algebra in
arbitrary dimensional space-time were reviewed, and their representations (reps) 
for the supermultiplets of all known supergravity theories were extensively searched by Strathdee.$^3$ 
This work has been an extremely useful guideline for studying supersymmetric theories. 
Cremmer$^4$ developed the complicated method for consistent trunctions 
by choosing a particular rep of real symplectic metric in order to derive 
{\it N}=6,4,2 supergravities from {\it N}=8 in five dimensions.
Recently, {\it M} and {\it F} theories$^5$ have been  also tackled from the point of view of the general properties of the superalgebra.$^6$

On the other hand, during last ten years, we have shown that superalgebras allow a more systematic analysis for finding the supermultiplets$^{7,8}$ of several supergravity and  superstring theories 
by using the Kac-Dynkin weight techniques of SU({\it m}/{\it n}) Lie superalgebra.$^9$
In particular, we have shown that the massless reps of supermultiplets of maximal supergravity theories$^{10,11}$ belong to only one irreduclble representation (irrep) of the SU(8/1) superalgebra.$^{12}$  
Recently, we have shown that all possible successive superalgebraic truncations 
from four-dimensional maximal {\it N}=8 supergravity theory 
to lower supersymmetric ones are systematically realised as 
sub-superalgebra chains of SU(8/1) superalgebra$^{13}$ 
by using projection matrices$^{14}$. 
Very recently, we have shown that the successive superalgebraic trunctions from 
{\it D}=10, {\it N}=2 chiral supergravity$^{10}$ to possible lower dimensional nonmaximal theories 
can be easily realized as sub-superalgebra chains of SU(8/1) Lie superalgebra.$^{15}$

In this paper, we show that the successive superalgebraic trunctions from the novel {\it D}=5, {\it N}=5 supergravity to possible lower dimensional theories can be systematically  realized as sub-superalgebra chains of SU(5/2) Lie superalgebra.
In Sec. II, we briefly recapitulate the mathematical structure of the SU(5/2) superalgebra related to 
{\it D}=5, {\it N}=5 supergravity. Through the introduction of atypical representations of SU(5/2) superalgebra, we newly find the supermultiplets of {\it D}=5, {\it N}=5 supergravity theory. 
Up to now, no one mentions the existence of one irrep of SU(5/2) supermultiplets
describing this theory.
In Sec. III, we explicitly show that supermultiplets of possible lower supersymmetric 
supergravity theories can be systematically obtained from SU(5/2) 
by successive superalgebraic dimensional reductions and truncations.
The last section contains conclusion.

\vspace{1cm}
\begin{center}
{\large \bf II. Kac-Dynkin Structure of SU(5/2) Superalgebra}
\end{center}

The Kac-Dynkin diagram of the SU(5/2) Lie superalgebra is

\begin{eqnarray}
w_1~~~~~w_2~~~~ w_3~~~~ w_4~~~~~ w_5~~~~ w_6~
\nonumber \\
\bigcirc \!\!-\!\!\!-\!\!\!-\!\!\bigcirc \!\!-\!\!\!-\!\!\!-\!\!
\bigcirc \!\!-\!\!\!-\!\!\!-\!\!\bigcirc \!\!-\!\!\!-\!\!\!-\!\!
\bigotimes \!\!-\!\!\!-\!\!\!-\!\!\bigcirc 
\end{eqnarray}

\noindent
where the set $(w_1~w_2~w_3~w_4~w_5~w_6)$ determines the highest-highest weight vector of an irrep.$^{1,2}$  Each weight component $w_i~(i \neq 5)$ of the highest-highest weight vector should be a nonnegative integer, while $w_5$ could be any {\it complex} number.  The first four white nodes and the last node form SU(5) $\otimes$ SU(2) bosonic subalgebra,  where SU(2) is isomorphic to SO(3) describing 
the massless modes of five dimensional space-time symmetry. The grey node is responsible to U(1) supersymmetric generator.

The corresponding graded Cartan matrix is

\vspace{0.5cm}
\begin{equation}
\left [
\begin{array}{rrrrrr}
2  & -1 & 0  & 0  & 0  & 0 \\
-1 & 2  & -1 & 0  & 0  & 0 \\
0  & -1 & 2  & -1 & 0  & 0 \\
0  & 0  & -1 & 2  & -1 & 0 \\
0  & 0  & 0  & -1 & 0  & 1 \\
0  & 0  & 0  & 0  & -1 & 2 \\

\end{array}
\right ]
 .
\end{equation}
\vspace{0.5cm}

\noindent
Let the positive and negative simple even roots of SU(5) $\otimes$ SU(2) bosonic subalgebra be $\alpha_i^{\pm} (i=1,2,3,4,6)$, and let the positive and negative simple odd roots be $\beta_5^{5\pm}$.  Other odd roots are easily obtained by the commutation relations such as

\begin{equation}
\beta_5^{i\pm} = [\alpha_i^\pm ,~ \beta_5 ^{i+1 \pm} ],~~~
\beta_6^{i\pm} = [\beta_5 ^{i \pm} ,~ \alpha_6^\pm],
~~~i=1,~2,~3,~4.
\end{equation}

\noindent
Then the action by an odd root $\beta_j^{i \pm}$ alternates a bosonic (fermionic) floor with a fermionic (bosonic) one.

The fundamental rep of SU(5/2) is (1~0~0~0~0~0), and it has the substructure of  $[({\bf 5,1,2})_B ~\oplus~({\bf 1,2,5})_F]$ in the 
basis of the SU(5)$\otimes$SU(2)$\otimes$U(1) 
bosonic subalgebra, where the subscripts {\it B} and {\it F} stand for bosonic and fermionic degrees of freedom, respectively, as follows:
\\
\begin{equation}
\begin{array}{lccl}
                 & (1~0~0~0~0~0) \\
\\
\mid \mbox{ground}> & (1~0~0~0~0~0) & = & ({\bf 5,1,2})_B \\
                 & \Downarrow \beta_5^{1-} \\
\mid \mbox{1st}> & (0~0~0~0~1~1) & = & ({\bf 1,2,5})_F. \\
\end{array}
\end{equation}
\\
\noindent
The U(1) supercharge generator is {\it Diag}(2,2,2,2,2,5,5) to satisfy the supertraceless condition. The complex conjugate rep of the fundamental rep is $(0~0~0~0~0~1) =$
$[({\bf 1,2,-5})_F~ \oplus$ $({\bf \overline {5},1,-2})_B]$ such as
\\
\begin{equation}
\begin{array}{lccl}
                 & (0~0~0~0~0~1) \\
\\
\mid \mbox{ground}> & (0~0~0~0~0~1) & = & ({\bf 1,2,-5})_F \\
                 & \Downarrow \beta^{5-}_6 \\
\mid \mbox{1st}> & (0~0~0~1~-1~0) & = & ({\bf \overline {5},1,-2})_B. \\
\end{array}
\end{equation}
\\

The even and odd roots consist of the adjoint rep $(1~0~0~0~0~1)$, 
which is obtained by the tensor product of the reps in Eqs.(4) and (5), 
\\
\begin{equation}
(1~0~0~0~0~0) \otimes (0~0~0~0~0~1) = (1~0~0~0~0~1) \oplus (0~0~0~0~0~0),
\end{equation}
\\
as follows

\begin{equation}
\begin{array}{llcl}
                 & (1~0~0~0~0~1) & \\
\\
\mid \mbox{gnd}> & (1~0~0~0~0~1) & = & \beta_j^{i+} \\
\\
\mid \mbox{1st}> & (1~0~0~1~0~0) & = & \mbox{SU(5)} \\
                 & (0~0~0~0~0~2) & = & \mbox{SU(2)} \\
                 & (0~0~0~0~0~0) & = & \mbox{U(1)} \\
\\
\mid \mbox{2nd}> & (0~0~0~1~0~1) & = & \beta_j^{i-}. \\
\end{array}
\end{equation}
\\

In general, the irreps of SU({\it m}/{\it n}) are divided into two types, which are {\it typical} and {\it atypical}.$^{1,9}$  All atypical reps of SU(5/2) are characterized by the fifth weight component $w_5$ of the highest-highest weight.  The atypicality condition$^9$ is given by

\begin{equation}
w_5 = \sum_{k=6}^{j} w_k -\sum_{k=i}^{4} w_k -10 + i +j,~~~1\leq i\leq 5,~~~ 5\leq j \leq 6.
\label{line1}
\end{equation}

\noindent
An atypical rep is obtained by terminating some odd root strings in a full weight system when $w_5$ satisfies the relation in Eq.(8) for specific {\it i}'s and {\it j}'s.  Thus the atypical reps generally have not equal bosonic and fermionic degrees of freedom.

On the other hand, all the typical reps of SU(5/2) consist of eleven floors, and have equal bosonic and fermionic degrees of freedom.  The lowest dimensional typical rep is
$(0~0~0~0~w_5~0)~=$ $[{\bf 512}_B \oplus {\bf 512}_F]$ for $w_5\neq -4,-3,-2,-1,0,1$.  
We take $w_5=-{3 \over 2}$ to make a real rep and normalize the U(1) supercharges by 3 such as

\begin{equation}
\begin{array}{lcc}
\mbox{floor}     & \mbox{SU(5/2)} & \mbox{SU(5) $\otimes$ SU(2) $\otimes$ U(1)}  \\
\noalign{\vskip3pt}
\noalign{\hrule}
\noalign{\vskip3pt}
\\
\mid \mbox{gnd}> & (0~0~0~0~-{3 \over 2} ~0)  & (0~0~0~0)(0)(-5) \\
\\
\mid \mbox{1st}> & (0~0~0~1~-{3 \over 2} ~1)  & (0~0~0~1)(1)(-4)  \\
\\
\mid \mbox{2nd}> & (0~0~1~0~-{1 \over 2} ~2)  & (0~0~1~0)(2)(-3) \\
                 & (0~0~0~2~-{5 \over 2} ~0)  & (0~0~0~2)(0)(-3) \\
\\
\mid \mbox{3rd}> & (0~1~0~0~+{1 \over 2} ~3)  & (0~1~0~0)(3)(-2) \\
                 & (0~0~1~1~-{3 \over 2} ~1)  & (0~0~1~1)(1)(-2) \\
\\
\mid \mbox{4th}> & (1~0~0~0~+{3 \over 2} ~4)  & (1~0~0~0)(4)(-1) \\             
                 & (0~1~0~1~-{1 \over 2} ~2)  & (0~1~0~1)(2)(-1) \\
                 & (0~0~2~0~-{3 \over 2} ~0)  & (0~0~2~0)(0)(-1) \\
\\
\mid \mbox{5th}> & (0~0~0~0~+{5 \over 2} ~5)  & (0~0~0~0)(5)(0) \\
                 & (1~0~0~1~+{1 \over 2} ~3)  & (1~0~0~1)(3)(0) \\
                 & (0~1~1~0~-{1 \over 2} ~1)  & (0~1~1~0)(1)(0) \\
\\
\mid \mbox{6th}> & (0~0~0~1~+{3 \over 2} ~4)   & (0~0~0~1)(4)(+1)  \\
                 & (1~0~1~0~+{1 \over 2} ~2)  & (1~0~1~0)(2)(+1) \\
                 & (0~2~0~0~-{1 \over 2} ~0)  & (0~2~0~0)(0)(+1) \\
\\
\mid \mbox{7th}> & (0~0~1~0~+{3 \over 2} ~3)  & (0~0~1~0)(3)(+2) \\
                 & (1~1~0~0~+{1 \over 2} ~1)  & (1~1~0~0)(1)(+2) \\
\\
\mid \mbox{8th}> & (0~1~0~0~+{3 \over 2} ~2)  & (0~1~0~0)(2)(+3) \\
                 & (2~0~0~0~+{1 \over 2} ~0)  & (2~0~0~0)(0)(+3) \\
\\
\mid \mbox{9th}> & (1~0~0~0~+{3 \over 2} ~1)  & (1~0~0~0)(1)(+4)\\
\\
\mid \mbox{10th}> & (0~0~0~0~+{3 \over 2} ~0)  & (0~0~0~0)(0)(+5). \\

\end{array}
\end{equation}
\\

As you know, since the maximum dimension of supermultiplets of the consistent supergravity theories is $[{\bf 128}_B \oplus {\bf 128}_F]$, it is impossible to accommodate the supermultiplets of five-dimensional theory in terms of the typical irrep.  Thus we should search atypical cases having lower dimensions in contrast to the typical ones, and have found the atypical reps of the type $(0~0~0~0~0~w_6)$ for $w_6 \geq 4$ are $[8(2w_6 -3)_B \oplus 8(2w_6 -3)_F]$ such as

\begin{equation}
\begin{array}{lccc}
\mbox{floor}     & \mbox{SU(5/2)} & \mbox{SU(5)$\otimes$SU(2)} & \mbox{dimension} \\
\noalign{\vskip3pt}
\noalign{\hrule}
\noalign{\vskip3pt}
\\
\mid \mbox{ground}> & (0~0~0~0~0~w_6)  & (0~0~0~0)(w_6) & w_6 +1 \\
\\
\mid \mbox{1st}>    & (0~0~0~1~-1~w_6 -1)  & (0~0~0~1)(w_6 -1) & 4w_6 \\
\\
\mid \mbox{2nd}>    & (0~0~1~0~-1~w_6 -2)  & (0~0~1~0)(w_6 -2) & 6(w_6 -1) \\
\\
\mid \mbox{3rd}>    & (0~1~0~0~-1~w_6 -3)  & (0~1~0~0)(w_6 -3) & 10(w_6 -2)\\
\\
\mid \mbox{4th}>    & (1~0~0~0~-1~w_6 -4) & (1~0~0~0)(w_6 -4) &  5(w_6 -3) \\
\\
\mid \mbox{5th}>    & (0~0~0~0~-1~w_6 -5) & (0~0~0~0)(w_6 -5) &  w_6 -4 \\        
\end{array}
\end{equation}
\\
Taking $w_6 = 4$, we get one graviton field in Eq.(10), since $w_6$ denotes the SU(2) $\approx$ SO(3) weight component.  The $\beta^{1-}_6$ string is terminated so that the fifth floor is terminated from the weight system in Eq.(10).  In fact, the only atypical rep containing the desired novel $D=5, ~N=5$ supergravity multiplets is $(0~0~0~0~0~4)=[{\bf 40}_B \oplus {\bf 40}_F]$  such as
\\
\begin{equation}
\begin{array}{lccc}
\mbox{floor}     & \mbox{SU(5/2)} & \mbox{SU(5)$\otimes$SU(2)} & \mbox{field} \\
\noalign{\vskip3pt}
\noalign{\hrule}
\noalign{\vskip3pt}
\\
\mid \mbox{ground}> & (0~0~0~0~0~4)  & (0~0~0~0)(4) & e_{\mu}^a \\
\\
\mid \mbox{1st}>    & (0~0~0~1~-1~3)  & (0~0~0~1)(3) & 5\Psi_{\mu}  \\
\\
\mid \mbox{2nd}>    & (0~0~1~0~-1~2)  & (0~0~1~0)(2) & 10A_{\mu}\\
\\
\mid \mbox{3rd}>    & (0~1~0~0~-1~1)  & (0~1~0~0)(1) & 10\lambda \\
\\
\mid \mbox{4th}>    & (1~0~0~0~-1~0) & (1~0~0~0)(0) & 5\phi \\
        
\end{array}
\end{equation}
\\
\noindent
Although the reps $(0~0~0~0~0~w_6)$ for $w_6 \geq 5$ have equal bosonic an fermionic degrees of freedom, they contain higher spin states,
which are higher than spin-2 ones, leading to inconsistent theories. 
On the other hand, the atypical reps for $w_6 \leq 3$ are 
$(0~0~0~0~0~3)=[{\bf 25}_B \oplus {\bf 24}_F]$ containing a gravitino,  
$(0~0~0~0~0~2)=[{\bf 13}_B \oplus {\bf 10}_F]$ having a Yang-Mills field, 
and $(0~0~0~0~0~1)=[{\bf 5}_B \oplus {\bf 2}_F]$ including a matter field.
Note that since all these reps have larger bosonic degrees of freedom than fermionic ones, it is impossible to construct the {\it D}=5, {\it N}=5 Yang-Mills theory with any combination of these reps.

\vspace{1cm}
\begin{center}
{\large \bf III. Successive Possible Superalgebraic Truncations \\
from {\it N}=5 to {\it N}=4,3,2,1}
\end{center}

\vspace{1cm}
\begin{center}
{\large \bf 3.1 {\it D}=5, {\it N}=4 Reduction}
\end{center}

Now, let us consider the possible superalgebraic truncations from the $D=5,~N=5$ to $D=5,~N=4,3,2,1$ supergravities.  One must carefully remove extra gravitino multiplets in a consistent manner in order to generate the existence of the $N=4,3,2,1$ theories.   The massless modes of the supermultiplets are in the rep space of SU({\it N}) $\otimes$ SU(2) $\subset$ SU({\it N}/2) supersymmetry.  The branching rules SU(5/2) $\longrightarrow$ SU(4/2) $\longrightarrow$ SU(3/2) $\longrightarrow$ SU(2/2) $\longrightarrow$ SU(1/2) are systematically attained by the successive removing of the first nodes from the Kac-Dynkin diagrams. 

A branching rule of SU(5/2) $\longrightarrow$ SU(4/2) for the rep in Eq.(11)  is                  

\begin{equation}
(0~0~0~0~0~4) \longrightarrow (0~0~0~0~4) \oplus (0~0~0~0~3)
\end{equation}

Then, the rep $(0~0~0~0~4) = [{\bf 24}_B \oplus {\bf 24}_F]$ of SU(4/2) is just 
the well-known graviton multiplet of $D=5,~N=4$ supergravity$^{3}$ as follows

\begin{equation}
\begin{array}{lccc}
\mbox{floor}     & \mbox{SU(4/2)} & \mbox{SU(4)$\otimes$SU(2)} & \mbox{field} \\
\noalign{\vskip3pt}
\noalign{\hrule}
\noalign{\vskip3pt}
\\
\mid \mbox{ground}> & (0~0~0~0~4)  & (0~0~0)(4) & e_{\mu}^a \\
\\
\mid \mbox{1st}>    & (0~0~1~-1~3)  & (0~0~1)(3) & 4\Psi_{\mu}  \\
\\
\mid \mbox{2nd}>    & (0~1~0~-1~2)  & (0~1~0)(2) & 6A_{\mu}\\
\\
\mid \mbox{3rd}>    & (1~0~0~-1~1)  & (1~0~0)(1) & 4\lambda \\
\\
\mid \mbox{4th}>    & (0~0~0~-1~0) & (0~0~0)(0) & \phi \\
        
\end{array}
\end{equation}
\\

\noindent
Note that the remaining rep $(0~0~0~0~3)=[{\bf 16}_B + {\bf 16}_F]$ makes 
an extra gravitino multiplet, which should be removed for consistency in the $D=5,~N=4$ theory.  On the other hand, the atypical reps for 
$w_6 \leq 2$ are given by $(0~0~0~0~2)=[{\bf 9}_B \oplus {\bf 8}_F]$, 
and $(0~0~0~0~1)=[{\bf 4}_B \oplus {\bf 2}_F]$.

\vspace{1cm}
\begin{center}
{\large \bf 3.2 {\it D}=5, {\it N}=3 Reduction}
\end{center}

The massless modes of supermultiplets of $D=5,~N=3$ are described in the rep space of SU(3) $\otimes$ SU(2) $\subset$ SU(3/2) superalgebra.  The branching rule of SU(4/2) $\longrightarrow$ SU(3/2) is

\begin{equation}
(0~0~0~0~4) \longrightarrow (0~0~0~4) \oplus (0~0~0~3).
\end{equation}

Then, the rep $(0~0~0~4)=[{\bf 14}_B + {\bf 14}_F]$ of SU(3/2) is identified to the graviton multiplets of $D=5,~N=3$ supergravity as follows

\begin{equation}
\begin{array}{lccc}
\mbox{floor}     & \mbox{SU(3/2)} & \mbox{SU(3)$\otimes$SU(2)} & \mbox{field} \\
\noalign{\vskip3pt}
\noalign{\hrule}
\noalign{\vskip3pt}
\\
\mid \mbox{ground}> & (0~0~0~4)  & (0~0)(4) & e_{\mu}^a \\
\\
\mid \mbox{1st}>    & (0~1~-1~3)  & (0~1)(3) & 3\Psi_{\mu}  \\
\\
\mid \mbox{2nd}>    & (1~0~-1~2)  & (1~0)(2) & 3A_{\mu}\\
\\
\mid \mbox{3rd}>    & (0~0~-1~1)  & (0~0)(1) & \lambda \\
        
\end{array}
\end{equation}
\\

\noindent
The other rep $(0~0~0~3)=[{\bf 10}_B + {\bf 10}_F]$ makes an extra gravitino multiplet, which should be removed for consistency in the $D=5,~N=3$ theory.

On the other hand, in contrast to the previous cases, at this level we can introduce an interesting
atypical rep $(0~0~0~2)=[{\bf 6}_B + {\bf 6}_F]$ describing a desired pure Yang-Mills multiplets such as

\begin{equation}
\begin{array}{lccc}
\mbox{floor}     & \mbox{SU(3/2)} & \mbox{SU(3)$\otimes$SU(2)} & \mbox{field} \\
\noalign{\vskip3pt}
\noalign{\hrule}
\noalign{\vskip3pt}
\\
\mid \mbox{ground}> & (0~0~0~2)  & (0~0)(2) & A_{\mu}\\
\\
\mid \mbox{1st}>    & (0~1~-1~1)  & (0~1)(1) & 3\lambda \\
\\
\mid \mbox{2nd}>    & (1~0~-1~0)  & (1~0)(0) & 3\phi\\
        
\end{array}
\end{equation}

The rest atypical rep $(0~0~0~1)$ is given by $[{\bf 3}_B \oplus {\bf 2}_F]$.

\vspace{1cm}
\begin{center}
{\large \bf 3.3 {\it D}=5, {\it N}=2 Reduction}
\end{center}

The massless modes of supermultiplets of $D=5,~N=2$ are in SU(2) $\otimes$ SU(2) $\subset$ SU(2/2).  The branching rule of SU(3/2) $\longrightarrow$ SU(2/2) is 

\begin{equation}
(0~0~0~4) \longrightarrow (0~0~4) \oplus (0~0~3),~~~~~\\
\\
(0~0~0~2) \longrightarrow (0~0~2) \oplus (0~0~1).
\end{equation}

\noindent
Then, the reps $(0~0~4)=[{\bf 8}_B + {\bf 8}_F]$, 
$(0~0~2)=[{\bf 4}_B + {\bf 4}_F],$ and 
$(0~0~1)=[{\bf 2}_B + {\bf 2}_F]$ of SU(2/2) is graviton multiplets, 
Yang-Mills multiplets, and matter multiplets of $D=5,~N=2$ supergravity, 
respectively, as follows;

\begin{equation}
\begin{array}{lccc}
\mbox{floor}     & \mbox{SU(2/2)} & \mbox{SU(2)$\otimes$SU(2)} & \mbox{field} \\
\noalign{\vskip3pt}
\noalign{\hrule}
\noalign{\vskip3pt}
\\
\mid \mbox{ground}> & (0~0~4)  & (0)(4) & e_{\mu}^a \\
\\
\mid \mbox{1st}>    & (1~-1~3)  & (1)(3) & 2\Psi_{\mu}  \\
\\
\mid \mbox{2nd}>    & (0~-1~2)  & (0)(2) & A_{\mu}\\
\\
\end{array}
\end{equation}

\begin{equation}
\begin{array}{lccc}
\mbox{floor}     & \mbox{SU(2/2)} & \mbox{SU(2)$\otimes$SU(2)} & \mbox{field} \\
\noalign{\vskip3pt}
\noalign{\hrule}
\noalign{\vskip3pt}

\mid \mbox{ground}> & (0~0~2)  & (0)(2) &  A_{\mu}\\
\\
\mid \mbox{1st}>    & (1~-1~1)  & (1)(1) & 2\lambda \\
\\
\mid \mbox{2nd}>    & (0~-1~0)  & (0)(0) & \phi \\
\\
\end{array}
\end{equation}

\begin{equation}
\begin{array}{lccc}
\mbox{floor}     & \mbox{SU(2/2)} & \mbox{SU(2)$\otimes$SU(2)} & \mbox{field} \\
\noalign{\vskip3pt}
\noalign{\hrule}
\noalign{\vskip3pt}

\mid \mbox{ground}> & (0~0~1)  & (0)(1) & \lambda \\
\\
\mid \mbox{1st}>    & (1~-1~0)  & (1)(0) & 2\phi \\
\\
\end{array}
\end{equation}

\noindent
Note that the rep $(0~0~3)=[{\bf 6}_B + {\bf 6}_F]$ denotes an extra gravitino multiplet, which should be removed for consistency in the $D=5,~N=2$ theory.

\vspace{1cm}
\begin{center}
{\large \bf 3.4 {\it D}=5, {\it N}=1 Reduction}
\end{center}

The massless modes of supermultiplets of $D=5,~N=1$ are in U(1) $\otimes$ SU(2) $\subset$ SU(1/2) superalgebra.  The branching rule of SU(2/2) $\longrightarrow$ SU(1/2) is

\begin{equation}
(0~0~4) \longrightarrow (0~4) \oplus (0~3),\\
(0~0~2) \longrightarrow (0~2) \oplus (0~1),\\
(0~0~1) \longrightarrow (0~1) \oplus (0~0).
\end{equation}

Then, the rep $(0~4)=[{\bf 5}_B \oplus {\bf 4}_F]$ of SU(1/2) may play a role of the graviton multiplet of $D=5,~N=1$ supergravity due to the existence of graviton as follows

\begin{equation}
\begin{array}{lccc}
\mbox{floor}     & \mbox{SU(1/2)} & \mbox{SU(2)} & \mbox{field} \\
\noalign{\vskip3pt}
\noalign{\hrule}
\noalign{\vskip3pt}
\\
\mid \mbox{ground}> & (0~4)  & (4) & e_{\mu}^a \\
\\
\mid \mbox{1st}>    & (-1~3)  & (3) & \Psi_{\mu}  \\
\\
\end{array}
\end{equation}
\\
\noindent
However, since this rep has asymmetry between bosons and fermions, we should couple this graviton multiplet to the following asymmetric matter one 
$(0~1)=[{\bf 1}_B \oplus {\bf 2}_F]$:

\begin{equation}
\begin{array}{lccc}
\mbox{floor}     & \mbox{SU(1/2)} & \mbox{SU(2)} & \mbox{field} \\
\noalign{\vskip3pt}
\noalign{\hrule}
\noalign{\vskip3pt}
\\
\mid \mbox{ground}> & (0~1)  & (1) & \lambda \\
\\
\mid \mbox{1st}>    & (-1~0)  & (0) & \phi \\
                
\end{array}
\end{equation}
\\

\noindent
Then, the coupled reps $(0~4) \oplus (0~1)=[{\bf 6}_B \oplus {\bf 6}_F]$ have same bosonic and fermionic degrees of freedom which make consistent graviton multiplets.  
Note that the rest rep $(0~3)=[\bf{3}_B + \bf{4}_F]$ makes an extra gravitino multiplet, which should be removed for consistency in the $D=5,~N=1$ theory.
On the other hand, the contents of the atypical rep 
$(0~2)=[{\bf 3}_B \oplus {\bf 2}_F]$ are given by

\begin{equation}
\begin{array}{lccc}
\mbox{floor}     & \mbox{SU(1/2)} & \mbox{SU(2)} & \mbox{field} \\
\noalign{\vskip3pt}
\noalign{\hrule}
\noalign{\vskip3pt}
\\
\mid \mbox{ground}> & (0~2)  & (2) & A_{\mu} \\
\\
\mid \mbox{1st}>    & (-1~1)  & (1) & \lambda \\
                
\end{array}
\end{equation}
\\

\noindent
Note that in contrast to the {\it D}=5, {\it N}=5 case, we can construct the desired {\it D}=5, {\it N}=1 Yang-Mills theory by using the coupled reps $(0~2) \oplus (0~1)=[{\bf 4}_B \oplus {\bf 4}_F]$ because $(0~1)$ rep in Eq.(23) has larger fermionic degrees of freedom than bosonic ones.

\vspace{1cm}
\begin{center}
{\large \bf IV. Conclusion}
\end{center}

In conclusion, we have newly studied a novel supermultiplets of 
{\it D}=5, {\it N}=5 supergravity in the context of SU(5/2) superalgebra.  
We have obtained possible regular maximal branching patterns 
in terms of Kac-Dynkin weight techniques.  
Then, we have shown that the possible superalgebraic truncations from 
the {\it D}=5, {\it N}=5 supergravity theory to 
the {\it D}=5, {\it N}=4,3,2,1 theories can be 
systematically realized as sub-superalgebra chains of the SU(5/2) superalgebra.
As results, we have explicitly identified the supermultiplets of 
the possible relevant lower supersymmetric theories, which have been classified 
in terms of super-Poincar\'{e} algebra by Strathdee, with irreps of 
SU({\it N}/2) superalgebra by using the systematic superalgebraic truncation method. 
Finally, since several authors$^{16,17}$ have recently considered $D=5$ supergravity
theories$^{18}$ through the compactification of $M$ theory, we hope 
through further investigations that our superalgebraic branching method will 
provide a deeper understanding of the structure of the supersymmetric systems 
including the $M$ and $F$ theories.

\vspace{1cm}

\begin{center}
{\bf Acknowledgments}
\end{center}

The present study was supported by the Basic Science Research
Institute Program, Korea Research Foundation, Project No. 1998-015-D00074.

\newpage

\begin{center}
{\bf REFERENCES}
\end{center}

\begin{description}
\item{1.} V. Kac, Adv. in Math. {\bf 26}, 8 (1977); Commun. Math. Phys.
{\bf 53}, 31 (1977).
\item{2.} Y.A. Gol'fand and E.P. Likhtman, Pis'ma Zh. Eksp. Theor. Fiz.
 {\bf 13}, 452(1971)[JETP Lett. {\bf 13},323(1971)];  A. Neveu and J. H. Schwarz, 
Nucl. Phys. {\bf B31}, 86 (1971); P. Ramond, Phys. Rev.
 {\bf D3}, 2415 (1971); P. G. O. Freund and  I. Kaplansky, J. Math. Phys.
 {\bf 17}, 228 (1976); S. Deser and B. Zumino, Phys. Lett. {\bf 62B}, 335 (1976);
D. Z. Freedman, P. van Nieuwenhuizen, and S. Ferrara, Phys. Rev. {\bf D13},
3214 (1976); Y. Ne'eman, Phys. Lett. {\bf 81B}, 190 (1979); 
F. Iachello, Phys. Rev. Lett. {\bf 44}, 772 (1980); 
A.B. Balantekin, I. Bars, and F. Iachello, {\it ibid.} {\bf 47}, 19 (1981); 
P. van Nieuwenhuizen, Phys. Rep. {\bf 68}, 189 (1981); 
M. B. Green and J. H. Schwarz, Nucl. Phys. {\bf B198}, 474 (1982); 
J. P. Hurni and B. Morel, J. Math. Phys. {\bf 24}, 157 (1983).
\item{3.} J. Strathdee, Int. J. Mod. Phys. {\bf A2}, 173 (1987).
\item{4.} E. Cremmer, in {\it Superspace and Supergravity}, edited by
S. Hawking and M. Rocek, p.267 (Cambridge University Press, London, England, 1980).
\item{5.} J. Schwarz, Phys. Lett. {\bf B367}, 97 (1996); 
 E. Witten, Nucl. Phys. {\bf B460}, 335 (1995);
 C. Vafa, Nucl. Phys. {\bf B469}, 403 (1996); 
 J. Polchiski, "Tasi Lectures on D-branes", {\tt hep-th/9611050}.
 H.J. Boonstra, B. Peeters, and K. Skenderis,  Nucl. Phys. {\bf B533}, 127 (1998)
 J. Maldacena, Adv. Theor. Math. Phys. {\bf 2}, 231 (1998).
\item{6.} I. Bars, Phys. Rev. {\bf D54}, 5203 (1996).
\item{7.} C. H. Kim, K. Y. Kim, W. S. l'Yi, Y. Kim, and Y. J. Park, 
Mod. Phys. Lett. {\bf A3}, 1005 (1988); 
C. H. Kim, Y. J. Park, K. Y. Kim, Y. Kim, and W. S. l'Yi, 
Phys. Rev. {\bf D44}, 3169 (1991).
\item{8.} C. H. Kim, K. Y. Kim, Y. Kim, H. W. Lee, W. S. l'Yi, and
Y. J. Park, Phys. Rev. {\bf D40}, 1969 (1989).
\item{9.} C. H. Kim, K. Y. Kim, W. S. l'Yi, Y. Kim, and Y. J. Park,
J. Math. Phys. {\bf 27}, 2009 (1986).
\item{10.} M. B. Green and J. H. Schwarz, Phys. Lett. {\bf 122B}, 143 (1983).
\item{11.} J. H. Schwarz and P. C. West, Phys. Lett. {\bf 126B}, 301 (1983);
P. Howe and P. C. West, Nucl. Phys. {\bf B238}, 181 (1984).
\item{12.} C. H. Kim, K. Y. Kim, Y. Kim, and Y. J. Park, Phys.
Rev. {\bf D39}, 2967 (1989).
\item{13.} C. H. Kim, Y. J. Park, and Y. Kim, Mod. Phys. Lett. {\bf A10}, 1929 (1995).
\item{14.} R. Slansky, Phys. Rep. {\bf 79}, 1 (1981); C. H. Kim, Y. J. Park,
I. G. Koh, K. Y. Kim, and Y. Kim, Phys. Rev. {\bf D27}, 1932 (1983).
\item{15.} C. H. Kim and Y. J. Park, Mod. Phys. Lett. {\bf A12}, 851 (1997).
\item{16.} A.C. Cadavid, A. Ceresole, R. D'Auria, and S. Ferrara, 
Phys. Lett. {\bf B357}, 76 (1995).
\item{17.} S. Mizoguchi and N. Ohta, Phys. Lett. {\bf B441}, 123 (1998);
J. Ellis, Z. Lalak, and W. Pokorski, "Five-Dimensional Gauged Supergravity and 
Supersymmetry Breaking in $M$ Theory", {\tt hep-th/9811133}.
\item{18.} A.H. Chamseddine and H. Nicolai, Phys. Lett. {\bf B96}, 89 (1980);
M. G$\ddot {\rm  u}$naydin, G. Sierra, and P.K. Townsend, Nucl. Phys. {\bf B253}, 573 (1985);
G.W. Gibbons, G. Horowitz, and P.K. Townsend, Class. Quant. Grav. {\bf 12}, 297 (1995);
A. Fujii and R. Kemmoku, ''$D=5$ Simple Supergravity on $AdS_{2} \times S^{3}$", 
{\tt hep-th/9903231}.

\end{description}
\end{document}